\documentclass[dvips]{acta}
\usepackage{supertabular,lscape,epsfig}
\usepackage{rotating}
\usepackage{amssymb}
\usepackage{amsmath}
\usepackage{esint}
\usepackage{longtable}
\DeclareSymbolFont{ppa}{OT1}{ppl}{m}{it}
\DeclareMathSymbol{\vv}{\mathalpha}{ppa}{'166}

\DeclareSymbolFont{ppa}{OT1}{ppl}{m}{it}
\DeclareMathSymbol{\vv}{\mathalpha}{ppa}{'166}

\SetPages{0}{0}

\SetVol{58}{2008}

\begin{document}

\begin{Titlepage}
\Title{Light curves of symbiotic stars in massive photomeric surveys II: ~~~~~~~~~ S and D'-type systems}
\Author{M.~~G~r~o~m~a~d~z~k~i$^{1}$,~~J.~~M~i~k~o~{\l}~a~j~e~w~s~k~a$^{2}$,
I.~~S~o~s~z~y~{\'n}~s~k~i$^{3}$}
{$^1$Departamento de F{\'i}sica y Astronom{\'i}a, 
Universidad de Valpara{\'i}so, Av. Gran Breta\~na 1111,
Playa Ancha, Casilla 5030, Chile\\
e-mail: {\small mariusz.gromadzki@uv.cl}\\
$^2$N.Copernicus Astronomical Center, Polish Academy of Sciences,
ul. Bartycka 18, 00-716 Warszawa, Poland\\
e-mail: {\small mikolaj@camk.edu.pl}\\
$^3$Warsaw University Observatory,
Al. Ujazdowskie 4, 00-478 Warszawa, Poland\\
e-mail: {\small soszynski@astrouw.edu.pl}}

\Received{Month Day, Year}
\end{Titlepage}

\Abstract{We present results of period analysis of ASAS, MACHO and OGLE light curves of 79 symbiotic stars classified as S and D'-type. The light curves of 58 objects show variations with the orbital period. In case of 34 objects, orbital periods are estimated for the first time, what increases the number of symbiotic stars with known orbital periods by about 64\%. The light curves of 46 objects show, in addition to the long-term or/and orbital variations, short-term variations with time scales of 50-200 days most likely due to stellar pulsations of the cool giant component. We also report eclipse-like minima and outbursts present in many of the light curves.}
{stars: activity -- stars: binaries: symbiotic -- surveys}

\Section{Introduction}
Symbiotic stars are long-period interacting binary systems, in which an evolved red giant transfers material onto its much hotter companion, which in most systems is a white dwarf. Based on their near-IR characteristics, symbiotic stars divide into two main classes (Allen 1982) depending whether the colours are stellar (S-type) or indicate a thick dust shell (D-type).
The majority ($\approx$80\%) of catalogued systems are S-type and have near-IR colours consistent with cool stellar photosphere temperatures of $\approx 3500-4000$\,K. Most of them have orbital periods of $\approx 500-1000$ days ({\it
e.g.} Miko{\l}ajewska 2012). The near-IR colours of D-type systems indicate the presence of a dust shell which obscures the star and re-emits at longer wavelengths. Near-IR photometric monitoring has shown that these D-type systems have large amplitude variations and that they contain Mira variables with pulsation periods in the range 300--600 days; they are often called symbiotic Miras (Whitelock, 1987). Since they must accommodate the Mira with its dust shell, these D-type systems should have much longer orbital periods than the S-types, a few tens of years and more. The latest review of symbiotic Miras and a comparison with normal Miras can be found in Whitelock(2003). There is also small subclass of symbiotic binaries contain earlier type of giant (F, G and K). These objects are called yellow symbiotics. Some of them show dust emission, these are signed as D'-type (Allen 1982).

Light curves of symbiotic stars reflect the very complex behaviour of these systems.  They show high and low activity stages, flickering, nova-like outbursts originating from the hot component (S \& D types), eclipses, ellipsoidal variability connected with orbital motion (S-type), radial pulsations (all D-type and some S-type) and semi-regular variation of the cool component (S-type), long-term dust obscuration (mostly D-type) and other types of variability (Miko{\l}ajewska 2001).

In this paper we analyse the light curves of 79 galactic S-type and D'-type symbiotic stars in different bands. The light-curves were provided by massive photometry surveys such as ASAS, MACHO, and OGLE. In some cases AAVSO light curves are analysed. Similar analysis of light curves of D-type symbiotic binaries was done by Gromadzki \etal (2009).

\Section{Data}

Belczy{\'n}ski \etal (2000) listed coordinates for symbiotic stars, but many of these are not sufficiently accurate to identify the symbiotics unambiguously. Therefore, we first identified the 2MASS counterparts using the existing finding charts and the Aladin Java graphics interface running at the CDS in Strasbourg. This works well because symbiotic stars, which have the near-IR colours of late-type giants, are intrinsically bright in $JHK$. The 2MASS coordinates were then used to identify symbiotic stars in the OGLE, MACHO and ASAS databases.

In the ASAS database (Pojma\'nski 2002)
\footnote{official home page of ASAS project: {\it
http://www.astrouw.edu.pl/asas/}}, the data were found for 102 symbiotic stars. However, only for 69 systems the quality of light curve was good enough for period analysis. These comprise $V$-band photometry obtained between November 2000 and August 2009. There are two limitations connected with these data. 
The first one is related to the brightness of objects in $V$ filter. Stars brighter than 8.5 mag are saturated whereas the limiting magnitude of the photometric system is $\approx$15 mag. However, we analysed ASAS light curves of objects brighter than 14 mag, because $\sigma_{V}$ in these cases was better than 0.2 mag.
The second constraint is associated with the instrumental angular resolution.
The scale of instrument is of about 15.5 arcsec per pixel. Since the image FWHM is of about 1.4--1.6 pixels and the size of the aperture was 2 pixels two stars are well separated if the distance between them is above $\approx 0.5$ arcmin. This means that in dense galactic regions objects are often blended. Information about blended objects are given in Table 1.

The OGLE-II/III database (Udalski \etal 1997, 2003)\footnote{official home page of OGLE project: {\it http://ogle.astrouw.edu.pl/}} includes light curves for 13 S-type galactic symbiotic stars. These comprise $I$-band photometry obtained between 1997 and 2009. 
Three stars, Hen~2-289, AS~269 and V3929~Sgr, showed only linear trends of unknown nature.
The light curves of remaining 10 were good enough for period analysis.
 
The MACHO database (Alcock \etal 1992) \footnote{official home page of MACHO project: {\it http://wwwmacho.mcmaster.ca/}} contains observations for 13 S-type systems obtained between 1993 and 1999. 
Only 9 of these light curves were good enough for period analysis.
For V3929~Sgr there are a few points only, whereas SS73~129, V4018~Sgr and Hen~2-379 are saturated.
The photometry was made through non-standard blue ($B_{\rm M}$) and red ($R_{\rm M}$) filters.

\Section{Period analysis}

All light-curves were analysed using the program PERIOD\footnote{the source program is available on {\it http://www.starlink.rl.ac.uk/}} ver. 5.0, based on the modified Lomb-Scargle method (Press \& Rybicki, 1989).
If it was necessary, long-term trends were removed by subtracting a polynomial of appropriate order. Sudden jumps of brightness (outbursts or eclipses) were also removed from light curves. The resultant power spectra of our targets were compared with the power spectra of windows.  The periods were derived from the inverse of the maximum of the peak in the periodogram ($f_{max}^{-1}$), whereas their accuracy was estimated by calculating the half-size of a single frequency bin ($\Delta
f$), centred on the peak ($f_{c}$ is the centre of the peak) of the periodogram and then converted to period units ($\Delta P = f_{c}^{-2} \cdot \Delta f$).

The highest peak in a typical power spectrum corresponds to variations with periods of 300--1000 days presumably related to orbital motion. 
The other strong peaks are connected with annual aliases, second and third harmonics, long-term variation and a combination thereof.

In the case of 24 objects included in our sample the orbital period was previously known from spectroscopic and/or photometric studies. 
These orbital periods were usually derived from observations covering longer periods than the photometric data used in our study, so we adopted them as more accurate.
  
The power spectra of residual light curves, with removed orbital modulation and/or long-term variation, often reveal peaks corresponding to periods of 50--200 days which may reflect pulsations of the red giant.
The orbital modulation was removed form light curves by fitting high (7-11) order spline polynomial. Such approach gave better results than subtracting a sinusoid because in many cases the amplitude of the orbital modulation showed cycle to cycle changes. Examples of our power spectra are shown in Fig. 1.

\begin{figure}
\begin{center}
\includegraphics[angle=0,width=11.5cm]{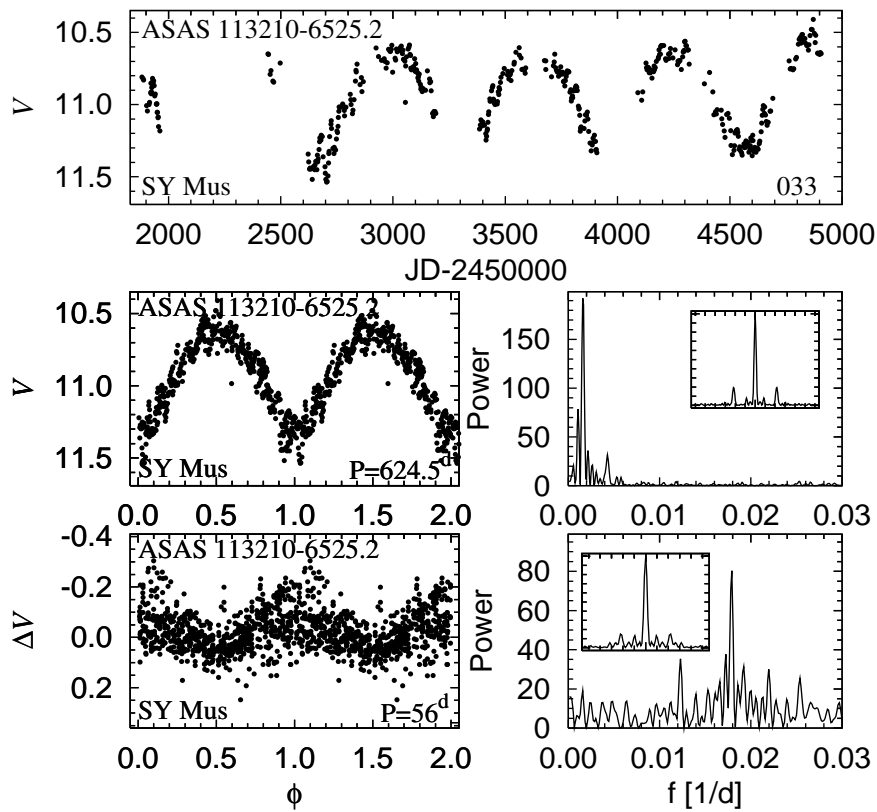}
\caption{ASAS light curve of SY Mus (top panel). ASAS light curve folded with orbital period (middle left panel) and related power spectra (middle right panel). ASAS light curve, after subtraction of orbital variations, folded with pulsation period (bottom left panels) and related power spectra (bottom right panels). Insights in top right corners of power spectra panels show power spectrum of windows.}
\end{center}
\end{figure}

\section{Results and discussion}

Results of our period analysis are summarized in Table 1.
The most important result of this study is detection of periodic light changes due to either orbital motion or pulsations or both.  The light curves folded with orbital periods are plotted in Figs. 2 and 3. Orbital ephemerides can be found in Table 2. In the case of 24 objects, we used more accurate orbital periods derived by other authors.
Residual light curves folded with pulsation periods are plotted in Figs. 4 and 5.

\begin{scriptsize}
\begin{longtable}{lcccccc}
\caption{Summary of periodicities derived from our analysis.}\\
\hline\hline 
No. & ~~~~~~Name~~~~~~ & ~~~~~~~~~~~~Survey~name~~~~~~~~~~~~  & ~~~~~~~~~$P_{\rm orb}$~~~~~~~~~ & ~~~~~~$P_{\rm pul}$~~~~~~ & ~~~~~~Other~periods~~~~~~~ & ~~~~~~Remarks~~~~~~ \\
\hline\hline 
\endfirsthead

\multicolumn{7}{c}%
{{\tablename\ \thetable{} -- continued}} \\
\hline\hline 
{No.} & {Name} & {Survey name} & {$P_{\rm orb}$}& {$P_{\rm pul}$} & {Other periods} & {Remarks}\\ 
\hline 
\endhead

\multicolumn{7}{c}{{Continued on next page}}\\ 
\hline\hline
\endfoot

\endlastfoot
012 & S 32         & ASAS 043745-0119.2   &              &            & & nc        \\
017 & V1261 Ori    & ASAS 052219-0840.0   & 628$\pm$24  &             & 640.5$^{[1]}$ & ecl, orb, pul \\
023 & BX~Mon       & ASAS 072523-0336.0   &              &            & 1401$^{[2]}$, 1259$^{[3]}$ & ecl, orb \\
024 & MWC~560      & ASAS 072551-0744.1   &              & 332$\pm$8  & 340, 1931$^{[4]}$& orb, pul \\
025 & Wray 15-157  & ASAS 080634-2832.0   &              &            & & bl \\
028 & AS 201       & ASAS 083143-2745.5   &              &            & & bl \\
031 & Hen 3-461    & ASAS 103909-5124.2   & 635$\pm$34   & 79$\pm$1   & & orb, pul \\
032 & SS73 29      & ASAS 110827-6547.3   &              &            & & fnt \\
033 & SY Mus       & ASAS 113210-6525.2   & 637$\pm$53   & 56$\pm$1   & 624.5$^{[5]}$ & ecl, orb, pul \\
035 & RT~Cru       & ASAS 123453-6433.9   & 325$\pm$9    & 63$\pm$1   & & orb, out, pul \\
039 & SS73 38      & ASAS 125126-6460.0   &              &            & & fnt \\
040 & Hen 3-863    & ASAS 130743-4800.3   & 1016$\pm$40  &            & & ecl, orb \\
042 & CD-36 8436   & ASAS 131602-3700.2   &              &            & & bl \\
043 & V840~Cen     & ASAS 132049-5550.2   & 792$\pm$26   &            & & orb \\
044 & Hen 3-905    & ASAS 133037-5758.3   &              &            & & bl \\
045 & RW Hya       & ASAS 133418-2522.8   & 375$\pm$12   &            & 370.3$^{[6]}$ &  ecl, orb \\
046 & Hen 3-916    & ASAS 133529-6445.7   & 803$\pm$54   & 150$\pm$2  & & bl, orb, pul \\
050 & V417 Cen     & ASAS 141559-6153.8   & 1652$\pm$11  &            & 245.7$^{[7]}$ & orb, pul? \\
051 & BD-21 3873   & ASAS 141634-2145.8   &   283$\pm$3  &            & 281.6$^{[8]}$ & orb \\
054 & Hen 3-1103   & ASAS 154828-4419.0   &  698$\pm$41  &            & & bl, orb \\
055 & HD 330036    & ASAS 155115-4845.0   & 1678$\pm$45  &            & & orb \\
057 & T CrB        & ASAS 155930+2555.2   &  225$\pm$3   &            & 227.6$^{[3]}$ & orb \\
059 & Wray 16-202  & ASAS 160657-4926.7   &              &            & & fnt \\
062 & QS Nor       & ASAS 162107-4223.9   & 244$\pm$5    &            & & orb, out \\
063 & Wray 15-1470 & ASAS 162321-2740.2   & 561$\pm$27   & 92$\pm$1   & & orb, pul \\
065 & Hen 3-1213   & ASAS 163515-5142.4   & 514$\pm$22   & 191$\pm$3  & & orb, pul \\
066 & Hen 2-173    & ASAS 163625-3951.7   &              &            & & fnt \\
068 & KX~TrA       & ASAS 164435-6237.2   &              & 178$\pm$3  & 1916$^{[9]}$& bl, out, pul \\
070 & HK~Sco       & ASAS 165441-3023.1   & 458$\pm$21   &            & & bl, orb, out \\
071 & CL~Sco       & ASAS 165452-3037.2   & 617$\pm$31   &            & 625$^{[10]}$ & orb, out \\
073 & V455~Sco     & ASAS 170722-3405.2   &              &            & 1398$^{[11]}$ & orb \\
074 & Hen 3-1341   & ASAS 170837-1726.5   & 626$\pm$26   & 105$\pm$2  & & orb, out, pul \\
075 & Hen 3-1342   & ASAS 170855-2323.6   & 562$\pm$28   & 44$\pm$1   & & bl, orb, pul \\
078 & Sa 3-43      & ASAS 171756-3001.7   &              &            & & fnt \\
086 & Th 3-30      & ASAS 173343-2807.4   &              &            & & fnt \\
088 & M 1-21       & ASAS 173417-1909.4   &              &            & 898$^{[11]}$ & fnt \\
093 & AE~Ara       & ASAS 174105-4703.3   & 771$\pm$87   & 126$\pm$2  & 803.4$^{[12]}$ & orb, out, pul \\ 
100 & Hen~2-289    & OGLE 174948.2-370127.0 &            &            & & nc \\
101 & RS Oph       & ASAS 175013-0642.5   & 464$\pm$25   &            & 453.6$^{[13]}$ & orb, out \\
104 & ALS 2        & ASAS 175051-1748.1   &              &            & & fnt \\
106 & Hen 2-294    & OGLE 175145.7-325451.5 & 393$\pm$15 & 49$\pm$1   & & orb, pul \\
    &              & ASAS 175145-3254.9   &              &            & &  fnt \\
107 & B1 3-14      & OGLE 175225.9-294557.0 &            & 160$\pm$2  & & pul \\
108 & B1 3-6       & OGLE 175256.3-311918.3 & 301$\pm$9  &            & & orb \\ 
109 & B1 L         & OGLE 17531378-3018056 &             &   86$\pm$1 & &  out, pul \\
112 & AS 255       & ASAS 175709-3515.6   &              &            & & bl \\
114 & H 2-34       & OGLE 175828.0-283341.9 & 459$\pm$20 & 109$\pm$1  & & orb, pul \\
115 & SS73 117     & ASAS 180223-3159.2   &              &            & &  bl \\
    &              & OGLE 180223.1-315912.7 &            & 179$\pm$5  & & ecl, pul       \\
116 & AS~269       & ASAS 180324-3242.4   &              &            & & bl\\
    &              & OGLE 180324.0-324224.9 &            &            & & nc \\
117 & Ap 1-8       & OGLE 180429.6-282134.7 &            &            & & orb \\
    &              & MACHO 109.20636.23   & 957$\pm$97   &   62$\pm$1 & & orb, pul \\
118 & SS73~122     & MACHO 101.20784.62   & 2409$\pm$10  & 189$\pm$4  & & orb, pul \\
    &              & OGLE 180441.2-270912.4 &            & 195$\pm$4  & & pul \\
119 & AS~270       & ASAS 180534-2020.6   & 794$\pm$61   &            & 671$^{[10]}$ & orb, out \\
121 & SS73~129     & ASAS 180706-2936.4   & 536$\pm$27   &            & & orb \\
    &              & MACHO 120.21787.8    &              &            & & sat \\
122 & Hen 3-1591   & ASAS 180732-2553.7   & 2350$\pm$41  &            & & bl, orb? \\
    &              & MACHO 179.21973.6    &              &            & &    \\
123 & V615~Sgr     & ASAS 180740-3606.3   & 657$\pm$39   & 70$\pm$1   & & orb, pul \\
124 & Ve 2-57      & ASAS 180822-2433.8   &              &            & & fnt \\
125 & AS 276       & ASAS 180910-4113.4   &              & 155$\pm$2  & & pul \\
127 & AS 281       & ASAS 181044-2757.9   &              &            & & bl, nc \\
    &              & MACHO 102.23372.26   & 533$\pm$30   & 65$\pm$1   & & orb, pul \\
128 & V2506 Sgr    & ASAS 181102-2832.7   &              &            & & bl \\
    &              & MACHO 110.23493.2498 & 868$\pm$80   & 93$\pm$1   & 913$^{[14]}$ & orb, pul \\
129 & SS73 141     & ASAS 181211-3310.7   &              &            & & out \\
130 & V343 Ser     & ASAS 181222-1140.1   & 511$\pm$22   &            & 450.5$^{[15]}$ & orb \\ 
131 & Y~CrA        & ASAS 181423-4250.5   &              & 84$\pm$1   & 1619$^{[12]}$ & bl, orb, pul \\
132 & YY~Her       & ASAS 181434+2059.3   & 580$\pm$42   &            & 589.5$^{[16]}$ & orb, out \\
133 & V2756 Sgr    & ASAS 181434-2949.4   & 480$\pm$19   &            & 243$^{[17]}$ & orb \\
134 & FG Ser       & ASAS 181507-0018.8   & 649$\pm$36   & 50$\pm$1   & 633.5$^{[18]}$ & ecl, orb, pul \\
135 & HD 319167    & ASAS 181525-3032.0   & 1744$\pm$37  &            & & orb \\
    &              & MACHO 123.25283.1    &              &            & & 100d \\
136 & Hen 2-374    & MACHO 305.35744.0    & 459$\pm$33   & 54$\pm$1   & 820$^{[12]}$ & orb, pul \\
137 & Hen 2-376    & MACHO 107.25453.25   &              &            & & 100d     \\
138 & V4074~Sgr    & ASAS 181605-3051.2   &              & 118$\pm$2  & &  pul \\
139 & V2905~Sgr    & ASAS 181720-2810.0   & 508$\pm$9    &            & & orb, out \\
140 & StHA 149     & ASAS 181856+2726.3   &              & 64$\pm$2   & & pul \\
141 & Hen 3-1674   & ASAS 182019-2622.6   &              &            & & bl \\
    &              & MACHO 163.27426.43   & 1004$\pm$20  &            & 1003$^{[14]}$ & ecl, orb \\
142 & AR~Pav       & ASAS 182029-6604.8   & 631$\pm$35   &            & 604.5$^{[19]}$ & ecl, orb, out \\
143 & V3929~Sgr    & MACHO 169.27679.3050 &              &            & & fp \\
    &              & OGLE 182058.9-264825.5 &            &            & & nc \\
144 & V3804 Sgr    & ASAS 182129-3132.1   & 426$\pm$8    &            & & orb, out        \\
145 & V443 Her     & ASAS 182208+2327.3   & 626$\pm$48   & 59$\pm$1   & 599.4$^{[18]}$ & orb, pul\\
146 & V3811~Sgr    & OGLE 182329.0-215309.5 &            & 139$\pm$15 & & ecl, pul \\
    &              & ASAS 182329-2153.1   &              &            & & bl, ecl? \\
147 & V4018~Sgr    & ASAS 182527-2836.0   & 513$\pm$22   & 93$\pm$1   & & orb, pul \\
    &              & MACHO 137.29602.905  &              &            & & sat \\
148 & V3890~Sgr    & OGLE 183043.3-240108.9 &            & 106$\pm$3  & 103.8, 519.7$^{[20]}$ & pul\\
151 & AS~316       & ASAS 184233-2117.8   &              & 62$\pm$1   & & pul \\
153 & MWC~960      & ASAS 184756-2005.8   &              & 183$\pm$3  & & pul \\
155 & AS~327       & ASAS 185317-2423.0   & 823$\pm$55   &  83$\pm$1  & & orb, pul \\
156 & FN~Sgr       & ASAS 185355-1859.7   & 563$\pm$28   &            & 568.3$^{[21]}$&  ecl, orb, out \\
158 & CM~Aql       & ASAS 190335-0303.3   &              &            & 1058$^{[22]}$  & bl, fp \\
159 & V919~Sgr     & ASAS 190346-1659.9   &              & 125$\pm$2  & & out, pul \\
160 & V1413~Aql    & ASAS 190347+1626.5   & 477$\pm$28   &            & 434.1$^{[23]}$ & ecl, orb \\
161 & NSV~11776    & ASAS 190955-0247.6   & 1625$\pm$16  & 198$\pm$4  & & orb, pul \\
168 & StHA~164     & ASAS 192842-0603.9   &              &            & & fnt \\
170 & Hen~3-1761   & ASAS 194225-6807.7   & 559$\pm$27   & 63$\pm$1   & 562$^{[24]}$   & orb, pul \\
171 & QW~Sge       & ASAS 194550-1836.8   &              &            & 390.5$^{[25]}$ & bl \\
176 & PU~Vul       & ASAS 202114+2134.3   &              & 138$\pm$2  & 4900$^{[26]}$  & ecl, pul  \\
177 & LT~Del       & ASAS 203557+2011.5   &              &            & 476.0$^{[27]}$ & bl  \\
178 & StHA~180     & ASAS 203920-0517.3   & 1494$\pm$38  &            & & orb \\
182 & CD-43~14304  & ASAS 210006-4238.8   &              &            & 1442, 1448$^{[28]}$& ecl, orb, out \\
184 & StHA~190     & ASAS 214145+0243.9   &              &            & & bl \\
185 & AG~Peg       & ASAS 215102+1237.5   & 743$\pm$68   &  55$\pm$1  & 816.5, 818.2$^{[15]}$ & orb, pul \\
s04 & CD-28~3719   & ASAS 070109-2906.4   &              & 198$\pm$3  & & pul \\
s06 & ZZ~Cmi       & ASAS 072414+0853.9   &              & 106$\pm$1  & & pul \\
s07 & NQ~Gem       & ASAS 073155+2430.2   &              & 58$\pm$1   & & pul \\
s08 & Wray~16-51   & ASAS 093329-4634.8   &              &            & & bl \\
s09 & Hen~3-653    & ASAS 112533-5956.5   &              & 115$\pm$1  & & bl, pul \\
s11 & CD-27~8661   & ASAS 122434-2818.9   & 753$\pm$47   & 85$\pm$1   & 763.3$^{[29]}$ & orb, orb \\
s12 & AE~Cir       & ASAS 144451-6923.5   &              &            & 342$^{[30]}$ & bl \\
s14 & V345~Nor     & ASAS 160644-5202.5   &              &            & & nc \\
s15 & V934~Her     & ASAS 170634+2358.3   & 44.08$\pm$0.1? &            & & orb/pul? \\
s16 & Hen~3-1383   & ASAS 172031-3309.9   &              &            & & fnt \\
s17 & V503~Her     & ASAS 173641+2318.2   &              & 130$\pm$2  & & ecl, pul \\
s23 & AS~280       & ASAS 180953-3319.7   &              &            & & bl, nc \\
s24 & AS~288       & ASAS 181248-2821.0   &              &            & & bl, nc \\
s25 & Hen~2-379    & ASAS 181617-2704.5   &              &            & & bl, nc \\
    &              & MACHO 168.25725.476  &              &            & & sat \\
s27 & V850~Aql     & ASAS 192335+0038.0   &              &            & & fnt \\
s28 & Hen~2-442    & ASAS 193943+2629.5   &              &            & & fnt, bl \\
\hline\hline
\end{longtable}
\end{scriptsize}
\begin{footnotesize}
\noindent
{\bf Legenda:} 100d - light curve covers only 100 days, bl - object blended, fp - few points, nc - none conclusive, orb - light curve shows variations with orbital period, pul - light curve shows pulsations, sat - object saturated, fnt - object too faint, out - light curve shows outburst, ecl - eclipse-like minimum in light curve.\\

\noindent
{\bf References:}  [1] Jorissen \etal 1998, [2] Dumm \etal 1998, 
[3] Fekel \etal 2000a, [4] Gromadzki \etal 2007a, [5] Schmutz \etal 1994, [6] Schild \etal 1996, 
[7] Van Winckel \etal 1994, [8] Smith \etal 1997, [9] Marchiano \etal 2008, [10] Fekel \etal 2007,
[11] Fekel \etal 2008, [12] Fekel \etal 2010, [13] Brandi \etal 2009, [14] Lutz \etal 2010, [15] Fekel \etal 2001,
[16] Miko{\l}ajewska \etal 2002, [17] Hoffleit 1970, [18] Fekel \etal 2000b, [19] Schild \etal 2001,
[20] Schaefer 2009, [21] Brandi \etal 2005, [22] Munari \etal 2001a, [23] Munari 1992,
[24] Brandi \etal 2006, [25] Munari \& Jurdana-Sepi{\'c} 2002, [26] Nussbaumer \& Vogel 1996, [27] Arkhipova, \etal 2011, [28] Schmid \etal 1998,
[29] Van Eck \etal 2000, [30] Mennickent \etal 2008.
\end{footnotesize}

\begin{table}[hpt]
\caption{Orbital ephemerides (references the same as in Table~1).}	
\begin{center}
\begin{scriptsize}
\begin{tabular}{cccc}\hline\hline
No. & Name    & Ephemeris & Reference\\
\hline
017 & V1261~Ori    & ${\rm Min}(V)=2\,451\,990 + 640.5\times E$    & [1] \\
023 & BX~Mon       & ${\rm Min}(V) =2\,449\,796 + 1259\times E$    & [3] \\
024 & MWC~560      & ${\rm Max}(V) =2\,448\,080 + 1931\times E$    & [4] \\
031 & Hen~3-461    & ${\rm Min}(V)=2\,452\,063 + 635\times E$      &   \\
033 & SY~Mus       & ${\rm Min}(V)=2\,452\,054 + 624.5\times E$    & [5] \\
035 & RT~Cru       & ${\rm Min}(V)=2\,452\,034 + 325\times E$      &   \\
040 & Hen~3-863    & ${\rm Min}(V)=2\,451\,721 + 1016\times E$     &   \\
043 & V840~Cen     & ${\rm Min}(V)=2\,452\,061 + 792\times E$      &   \\
045 & RW~Hya       & ${\rm Min}(V)=2\,451\,738 + 370.3\times E$    & [6] \\
046 & Hen~3-916    & ${\rm Min}(V)=2\,452\,410 + 803\times E$      &   \\
050 & V417~Cen     & ${\rm Min}(V)=2\,452\,613 + 1652\times E$     &   \\
051 & BD-213873    & ${\rm Min}(V)=2\,451\,863 + 281.6\times E$    & [8] \\
054 & Hen~3-1103   & ${\rm Max}(V)=2\,452\,211 + 698\times E$      &   \\
055 & HD~330036    & ${\rm Min}(V)=2\,451\,048 + 1678\times E$     &   \\
057 & T~CrB        & ${\rm Min}(V)=2\,447\,919 + 227.6\times E$    & [3] \\
062 & QS~Nor       & ${\rm Min}(V)=2\,452\,024 + 244\times E$      &   \\
063 & Wray~15-1470 & ${\rm Min}(V)=2\,451\,845 + 561\times E$      &   \\
065 & Hen~3-1213   & ${\rm Min}(V)=2\,451\,806 + 514\times E$      &   \\
070 & HK~Sco       & ${\rm Min}(V)=2\,452\,023 + 458\times E$      &   \\
071 & CL~Sco       & ${\rm Min}(V)=2\,452\,018 + 625\times E$      & [10] \\
073 & V445 Sco     & ${\rm Min}(V)=2\,452\,641.5 + 1398\times E$   & [11] \\
074 & Hen~3-1341   & ${\rm Min}(V)=2\,451\,970 + 626\times E$      &   \\
075 & Hen~3-1342   & ${\rm Min}(V)=2\,452\,287 + 562\times E$      &   \\
093 & AE~Ara       & ${\rm Min}(V)=2\,453\,449 + 803.4\times E$      & [12] \\
101 & RS Oph       & ${\rm Min}(V)=2\,451\,848 + 453.6\times E$    & [13] \\
106 & Hen~2-294    & ${\rm Min}(I)=2\,451\,961 + 393\times E$      &   \\
108 & B1~3-6       & ${\rm Min}(I)=2\,451\,902 + 301\times E$      &   \\
114 & H~2-34       & ${\rm Min}(I)=2\,451\,974 + 459\times E$      &   \\
117 & Ap~1-8       & ${\rm Min}(B_{\rm M})=2\,448\,973 + 957\times E$  &   \\
118 & SS73~122     & ${\rm Min}(R_{\rm M}/I)=2\,446\,709 + 2409\times E$  &   \\
119 & AS~270       & ${\rm Min}(V)=2\,451\,633 + 671\times E$      & [10]  \\
121 & SS73~129     & ${\rm Min}(V)=2\,452\,220 + 536\times E$      &   \\
122 & Hen~3-1591   & ${\rm Min}(B_{\rm M}/V)=2\,451\,310 + 2350\times E$ & \\
123 & V615~Sgr     & ${\rm Min}(V)=2\,452\,168 + 657\times E$      &   \\
127 & AS281        & ${\rm Min}(B_{\rm M})=2\,449\,021 + 533\times E$  &   \\
128 & V2506~Sgr    & ${\rm Min}(B_{\rm M})=2\,448\,781 + 868\times E$  &   \\
130 & V343~Ser     & ${\rm Min}(V)=2\,450\,724.7 + 450.5\times E$  & [15] \\
131 & Y~CrA        & ${\rm Min}(V)=2\,454\,295 + 1619\times E$     & [12] \\
132 & YY~Her       & ${\rm Min}(V)=2\,450\,686.2 + 589.5\times E$  & [16] \\
133 & V2756~Sgr    & ${\rm Min}(V)=2\,451\,894 + 480\times E$      &    \\
134 & FG~Ser       & ${\rm Min}(V)=2\,451\,665 + 633.5\times E$    & [18] \\
135 & HD~319167    & ${\rm Min}(V)=2\,451\,756 + 1744\times E$     &  \\
136 & Hen~2-374    & ${\rm Min}(B_{M})=2\,452\,968 + 820\times E$  &  [12]  \\ 
139 & V2905~Sgr    & ${\rm Max}(V)=2\,451\,630 + 508\times E$      &   \\
141 & Hen~3-1674   & ${\rm Min}(B_{\rm M}/R_{\rm M})=2\,449\,178 + 1004\times E$ &   \\
142 & AR~Pav       & ${\rm Min}(V)=2\,448\,139 + 604.5\times E$    & [19] \\
144 & V3804~Sgr    & ${\rm Min}(V)=2\,451\,439 + 426\times E$      &   \\
145 & V443~Her     & ${\rm Min}(V)=2\,450\,197.3 + 599.4\times E$  & [18] \\
147 & V4018~Sgr    & ${\rm Min}(V)=2\,452\,129 + 513\times E$      &   \\
155 & AS~327       & ${\rm Min}(V)=2\,451\,954 + 823\times E$      &   \\
156 & FN~Sgr       & ${\rm Min}(V)=2\,450\,270 + 568.3\times E$    & [21] \\
160 & V1413~Aql    & ${\rm Min}(V)=2\,446\,650 + 434.1\times E$    & [23] \\
161 & NSV~11776    & ${\rm Min}(V)=2\,451\,672 + 1625\times E$     &   \\
170 & Hen~3-1761   & ${\rm Min}(V)=2\,451\,650 + 562\times E$      & [24]  \\
178 & StHA~180     & ${\rm Min}(V)=2\,451\,332 + 1494\times E$     &   \\
185 & AG~Peg       & ${\rm Min}(V)=2\,431\,667.5 + 816.5\times E$  & [15] \\
s11 & CD~27-8661   & ${\rm Min}(V)=2\,452\,169 + 763.3\times E$    & [29] \\ 
s17 & V503~Her     & ${\rm Min}(V)=2\,453\,145 + 1575\times E$     &    \\ 
\hline\hline
\end{tabular}
\end{scriptsize}
\end{center}
\label{tab:efm}
\end{table}

\clearpage
\begin{figure}
\begin{center}
\includegraphics[angle=0,width=11.5cm]{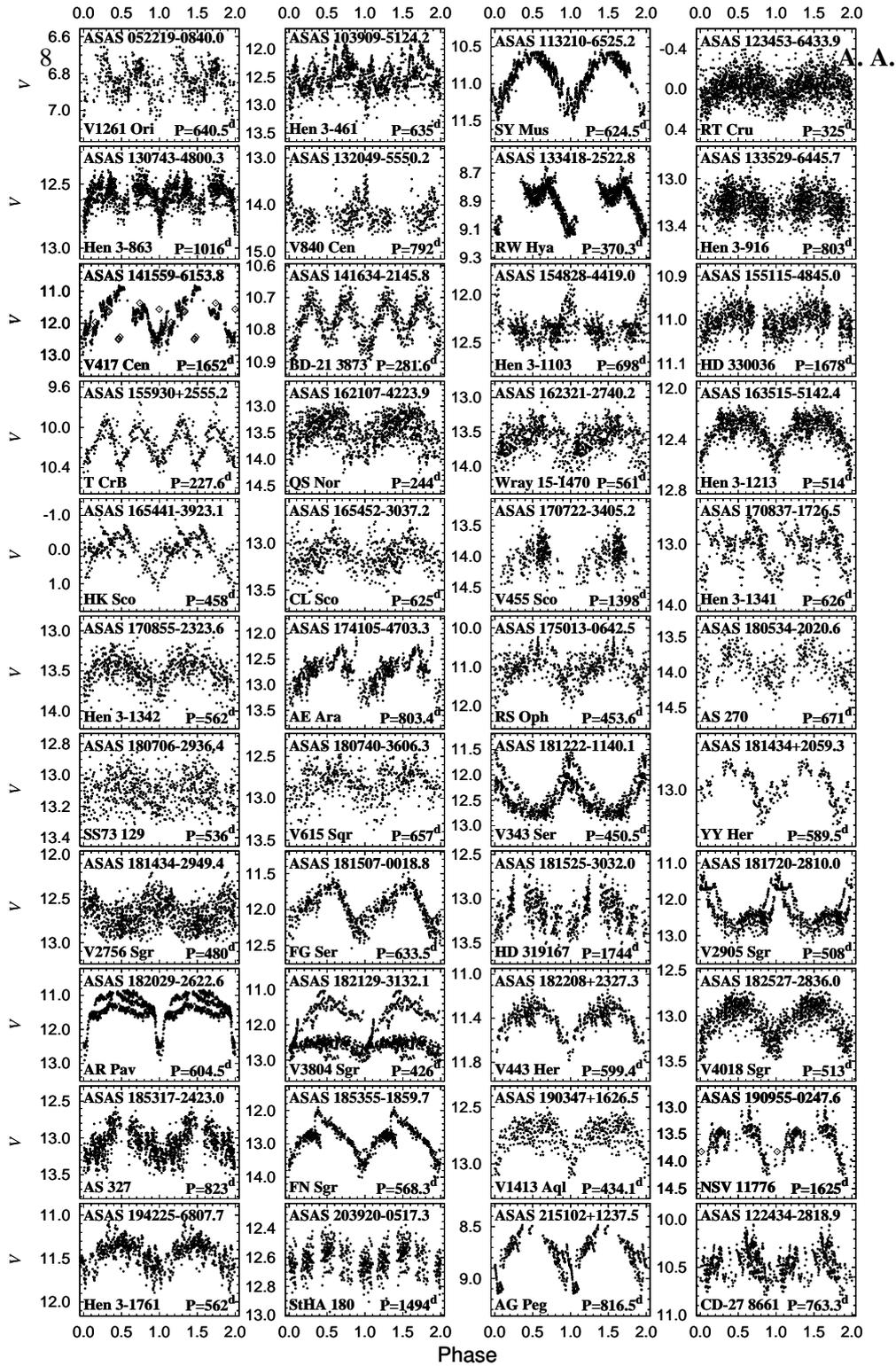}
\caption{ASAS light curves folded with orbital periods.}
\end{center}
\end{figure}

\clearpage
\begin{figure}
\begin{center}
\includegraphics[angle=0,width=11.5cm]{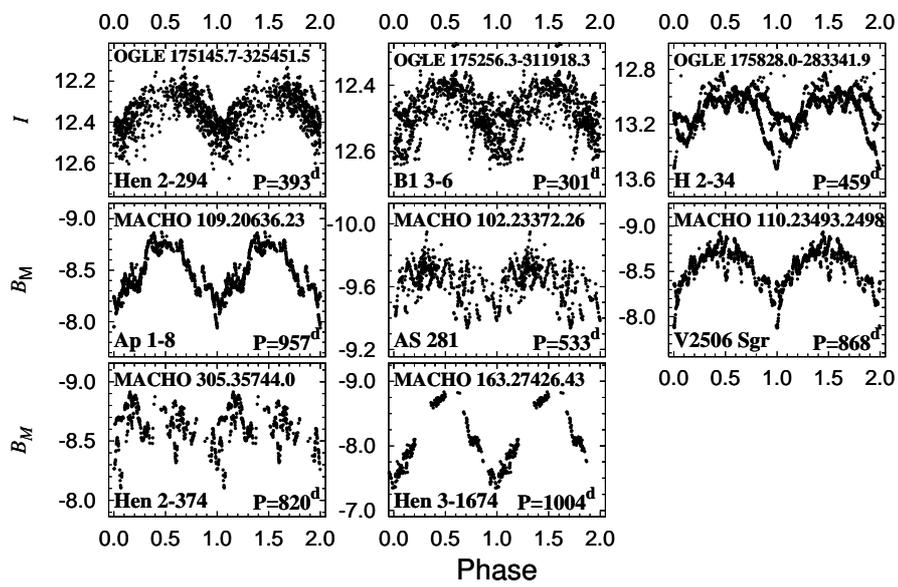}
\caption{OGLE and MACHO light curves folded with orbital periods.}
\end{center}
\end{figure}

\clearpage
\begin{figure}
\begin{center}
\includegraphics[angle=0,width=11.5cm]{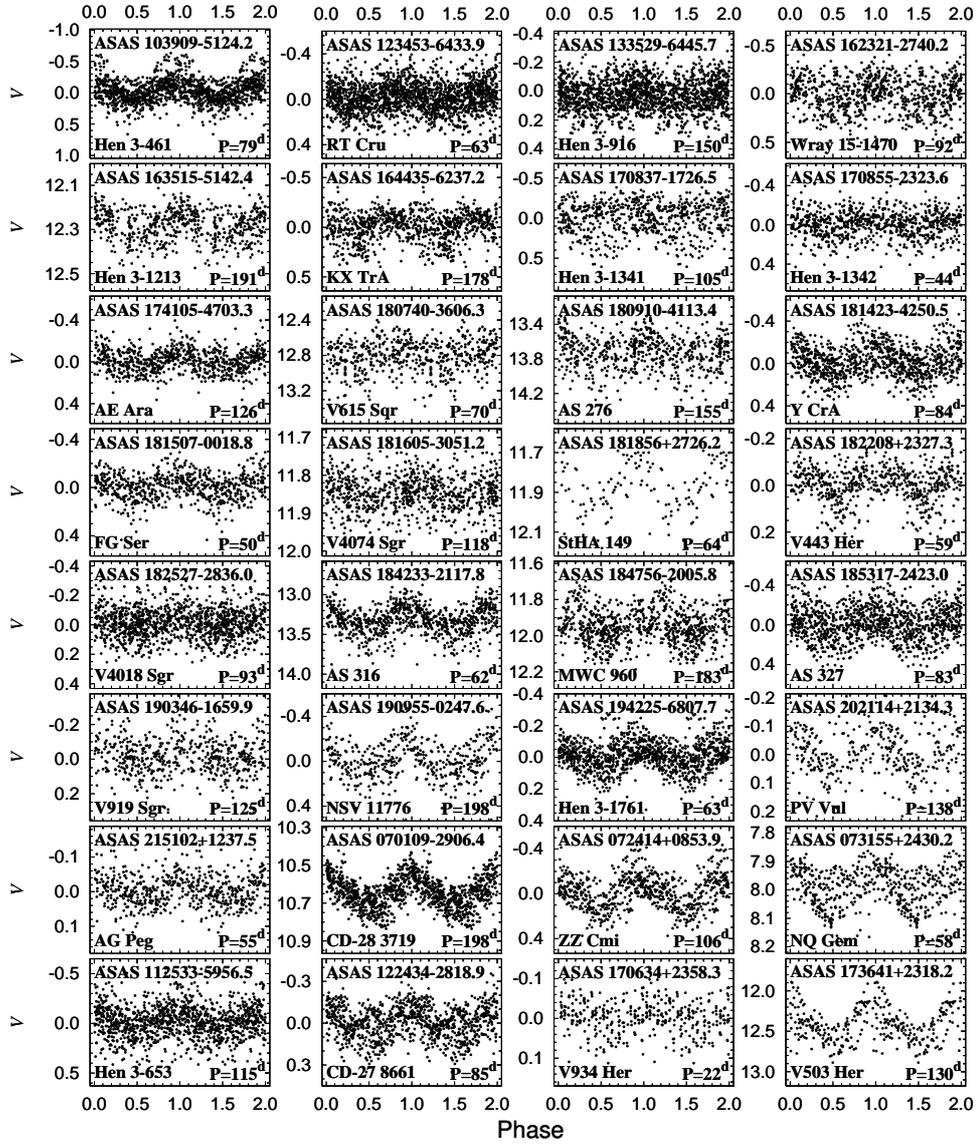}
\caption{ASAS light curves folded with pulsation periods. In most cases, orbital variations were subtracted.}
\end{center}
\end{figure}

\clearpage
\begin{figure}
\begin{center}
\includegraphics[angle=0,width=11.5cm]{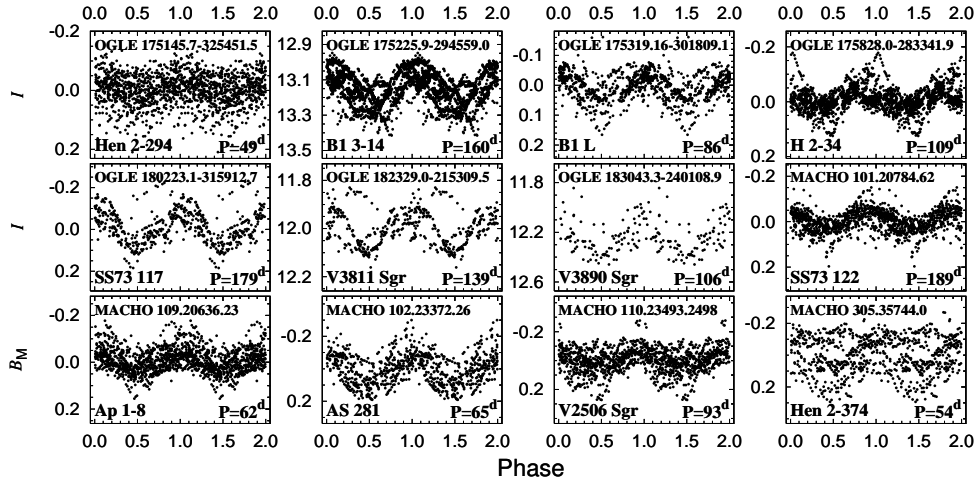}
\caption{OGLE and MACHO light curves folded with pulsation periods. In most cases, orbital variations were subtracted.}
\end{center}
\end{figure}

\subsection{Orbital periods}

The orbitally related light changes in symbiotic binaries can be caused by: (i) reflection effect, (ii) ellipsoidal variations, (iii) eclipses, and (iv) periodic brightening caused by increasing of accretion rate in eccentric systems during the periastron passage.
It is not obvious that long-period ($>$200 days) variations are caused by orbital motion. 
However, in favour of such an interpretation is fact that among 58 objects showing such changes, 24 systems have known orbital periods from previous photometric and spectroscopic studies. Most of these systems have also spectroscopic orbits determined from radial velocities of the cool component absorption features.

The main cause of orbital light curve modulation in our sample is reflection effect, observed in 37 light curves analysed for this study. However, in contrast to the classical case, in symbiotic stars the hot component radiation illuminates and partly ionizes the cold giant wind rather than its surface.

The light curves of several systems show more or less pronounced secondary minima, and their shape can be interpreted in terms of ellipsoidal changes in the red giant and variable nebular emission due to reflection effect. 
These are: V1261~Ori, Hen 3-863, RW~Hya, BD-21~3873, T~CrB, Hen~3-1341, Ap~1-8, 
Hen~2-374, YY~Her, V1413~Aql and V934~Her. The ellipsoidal variability in RW~Hya, BD-21~3873, T~CrB and YY~Her was reported and studied by different groups (Rutkowski \etal 2007, Smith \etal 1997, Belczy\'nski \& Miko{\l}ajewska 1998, and Miko{\l}ajewska \etal 2002, respectively) whereas in the remaining systems such variability has been detected for the first time. 
The ellipsoidal effect is dominating the V-band (ASAS) light curves of V1261~Ori, Hen 3-863, BD-21~3873, T~CrB, {\it i.e.} systems whose hot component have relatively low (as for a symbiotic star) luminosity, $\lesssim 100 \LS$ or so, and their optical spectra are dominated by the cool giant. In the case of systems with more luminous hot component, the secondary minimum is partly veiled by the illumination effect (like \eg RW~Hya) and even completely obscured. 
The ellipsoidal changes also vanish during optical outbursts when strong blue, A/F-type, spectrum completely veils the red giant features in the optical range (Miko{\l}ajewska \etal 2003).
For example, the near infrared light curves of RW~Hya, SY~Mus, and AR~Pav are evidently ellipsoidal (Rutkowski \etal 2007) whereas the V light cuves presented in this study show shallow secondary minimum only in RW~Hya. The V light curve of SY~Mus is dominated by illumination effect while in AR Pav a strong A/F-type shell is permanently present (Quiroga \etal 2002).
Similarly, the OGLE/I light curve of Ap~1-8 shows ellipsoidal modulation, while the MACHO/$B_{\rm M}$ light curve shows changes caused by reflection effect (Fig. 6).

\begin{figure}[htp]
\begin{center}
\includegraphics[width=11.5cm]{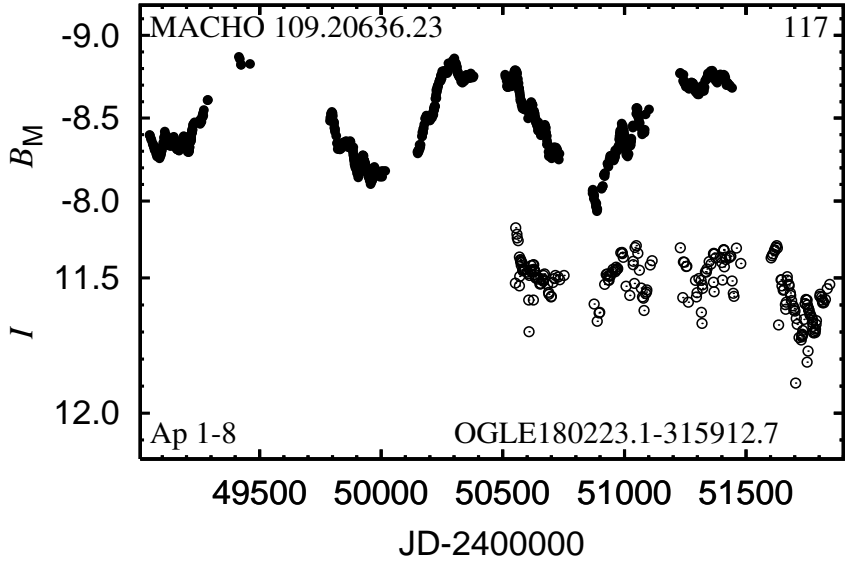}
\end{center}
\caption{OGLE and MACHO light curves of Ap~1-8.}
\end{figure}

Hen~3-1341 is very active. Its visual light curve covers an outburst associated with jets ejection (Munari \etal 2005) as well as quiescence state. Possible sinusoidal variations with a period of 626 days are visible during quiescence state (Fig. 7).

\begin{figure}[htp]
\begin{center}
\includegraphics[width=11.5cm]{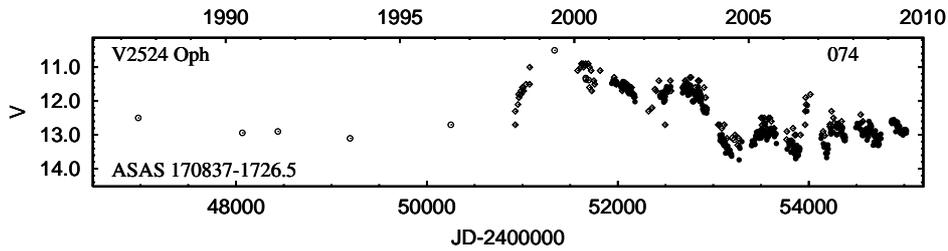}
\end{center}
\caption{The visual light curves of Hen 3-1341. Dots represent $V$-band ASAS data, open dots visual AAVSO observations and open diamonds are $V$-band measurements from Munari \etal (2005).}
\end{figure}

The ASAS light curve of YY~Her covers the decline from its last outburst. The primary minimum is shifted, while the secondary one is barely visible. Formiggini and Leibowitz (2006) showed that $P_{\rm orb}=593.2$ days modulates the quiescent light curve of YY~Her whereas a periodic oscillation with a shorter period of 551.4 days dominates the outburst light curve.
Such a secondary periodicities, always $\approx 10-20\, \%$ shorter than the orbital period, are often observed in the outburst light curves of many symbiotic stars, and the nature of this behaviour is poorly understood ({\it e.g.} Miko{\l}ajewska 2003). 

A weak secondary minimum may also be present in the ASAS light curve of V1413~Aql, and given its relatively short orbital period, it is very promising candidate to search for ellipsoidal changes at longer wavelengths.

V934~Her can be another possible ellipsoidal variable (it is one of the rare symbiotic systems hosting a neutron star). Although, it would require a relatively short orbital period (as for a symbiotic star), of $\approx$44 days only. The nature of this variability is not clear and it could be caused by either the red giant pulsations or reflection effect. One should also mention that Masetti \etal (2002) and Galloway \etal (2002) found periodicity of 400 days based on broad-band X-ray data, and optical radial velocities, respectively. However, this period was not confirmed by analysis of longer duration X-ray light curves (Corbet \etal 2008).
The ASAS light curves of V934~Her folded with periods of 44.08 and 22.04 days and corresponding power spectrum are plotted in Fig. 8.

\begin{figure}[hpt]
\begin{center}
\includegraphics[width=11.5cm]{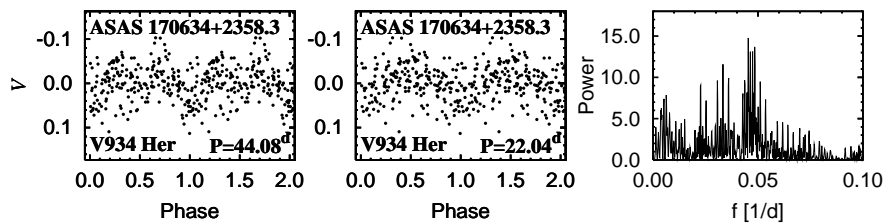}
\end{center}
\caption{The ASAS light curves of V934~Her folded with periods of 44.08 days and 22.04 days (left and middle panels, respectively) and the power spectrum (right panel).} 
\end{figure}

Light curves of 15 systems show one or more sharp and deep minima which may be caused by an eclipse. 
Among them, well known eclipsing systems ({\it e.g.} AR~Pav and PU~Vul) are present. 
In other cases the moment of minimum agrees fairly well with the time of spectroscopic conjunction ({\it e.g.} CD-43~14304).
In the case of three objects (SS73~117, V3811~Sgr and V503~Her) it is not clear whether they are really eclipsing because their light curves show only one or two minima. 
The situation is much better in the case of Hen 3-863 and Hen~3-1674. 
The ASAS light curve of Hen~3-863 shows three minima
(two primary and one secondary) and their overall shape is typical for an eclipsing binary. In the case of Hen~3-1674, the eclipse is confirmed by the spectrum available in the literature (Allen 1984, Medina Tanco \& Steiner 1995, and Munari \& Zwitter 2002). The eclipse of Hen~3-1674 is shown in Fig. 9. More examples of eclipses are shown in Fig. 10. The observed eclipses are summarized in Table 3. Eclipses of AS~269 and V4074~Sgr announced in Gromadzki \etal (2007b) have not been confirmed.

\begin{figure}[htp]
\begin{center}
\includegraphics[width=11.5cm]{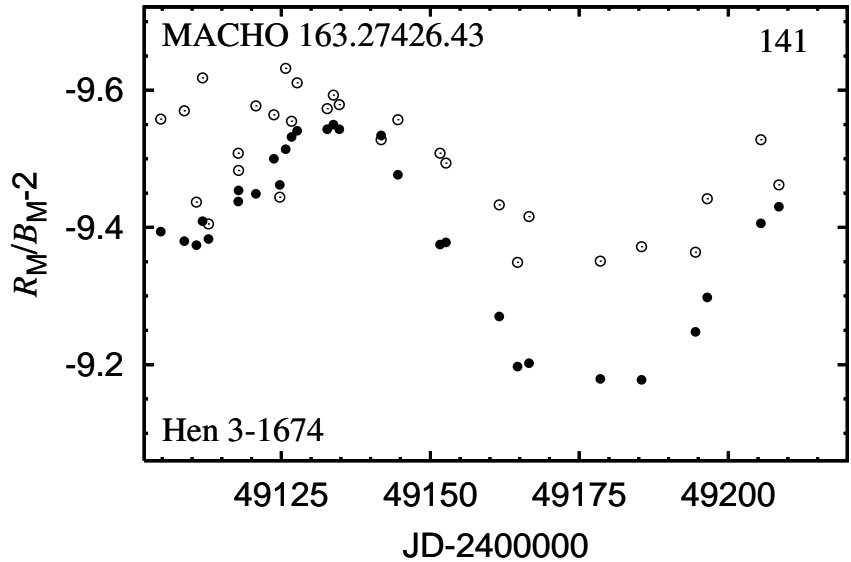}
\end{center}
\caption{Part of the MACHO light curve of Hen 3-1674 showing eclipse in July 1993 ($\approx$JD~2\,449\,175). Open dots represent the measurements obtained in $B_{\rm M}$ filter, and filled dots in $R_{\rm M}$ filter.}
\end{figure}

\begin{figure}[htp]
\begin{center}
\includegraphics[width=11.5cm]{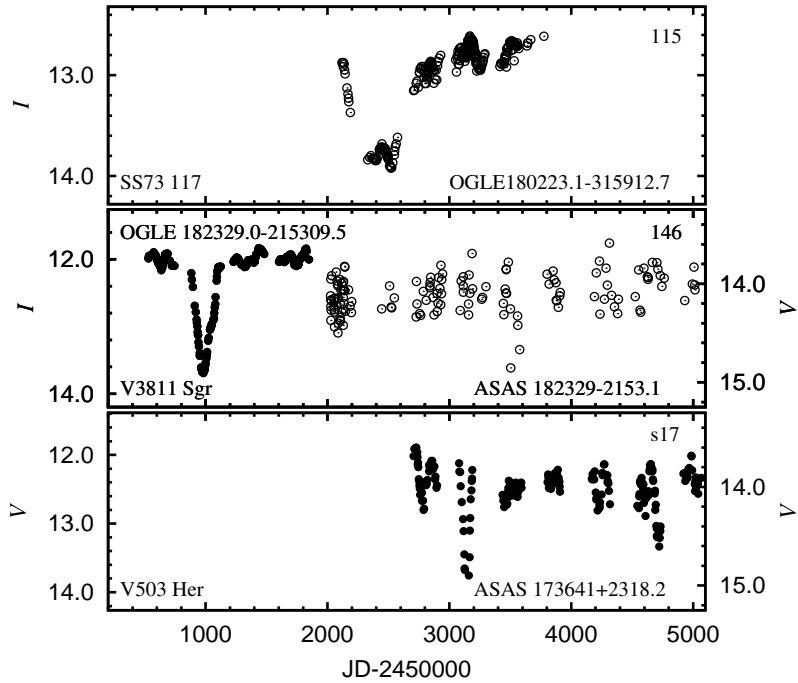}
\end{center}
\caption{Examples of light curves showing eclipses.}
\end{figure}

\begin{table}[hpt]
\caption{Symbiotic stars showing eclipses.}
\label{tab:zac}
\begin{center}
\begin{footnotesize}
\begin{tabular}{cccc}\hline\hline
No. & Name   & Observed minima (JD\,2\,400\,000$+$) \\
\hline
017 & V1261~Ori    & 54\,552\\
023 & BX~Mon       & 52\,022, 53\,281, 54\,540  \\
033 & SY~Mus       & 52\,679, 54\,552\\ 
040 & Hen~3-863    & 52\,737, 53\,753, 54\,261\\
045 & RW~Hya       & 52\,108, 52\,479, 52\,849, 53\,219, 53\,960, 54\,330, 54\,700\\
115 & SS73~117     & 52\,320 \\ 
134 & FG~Ser       & 52\,932, 53\,566, 54\,199\\
141 & Hen~3-1674   & 49\,178 \\
142 & AR~Pav       & 52\,975, 53\,580, 54\,184, 54\,789  \\
146 & V3811~Sgr    & 50\,980, 53\,540?  \\
156 & FN~Sgr       & 52\,543, 53\,112, 53\,680, 54\,248 \\
160 & V1413~Aql    & 52\,727, 53\,162, 53\,596\\
176 & PU~Vul       & 44\,550, 49\,450, 54\,350  \\
182 & CD-43~14304  & 52\,770, 53\,570, 54\,320  \\
s17 & V503~Her     &  53\,145, 54\,720  \\ 
\hline\hline
\end{tabular}
\end{footnotesize}
\end{center}
\end{table}

Two objects, BX~Mon and CD-43~14304, show periodic brightenings related to the periastron passage according to their known spectroscopic orbits (Fekel \etal 2000a, Schmid \etal 1998). Light curves of these objects are shown in Fig. 11.
The brightenings always happen a few hundred days after the periastron, and they are probably caused by enhanced accretion rate. In the case of BX~Mon spectroscopic observations have confirmed enhancements in the hot component activity following the periastron passage (Anupama \etal 2012). 
Such a behaviour is also observed in visual light curve of MWC~560 ({\it e.g.} Gromadzki \etal 2007a). 
We think that the same effect can be responsible for periodic 'outbursts' present in light curves of V840~Cen, Hen~3-1103, 
and V2905 Sgr, although the orbital periods for these systems are shorter, of 500-800 days (see Fig. 2).
In case of KX~TrA, outburst in 2003 was also preceded by periastron passage (according orbital solution derived by Marchiano \etal 2008).
However, AAVSO light curve of this object did not show brightenings after previous periastron passages (see Fig. 2 in Marchiano \etal 2008), what means last brightening had different nature that these in BX~Mon or CD-43~14304 and it was most likely 
nova-like outburst typical for classical symbiotic stars.

\begin{figure}[htp]
\begin{center}
\includegraphics[width=11.5cm]{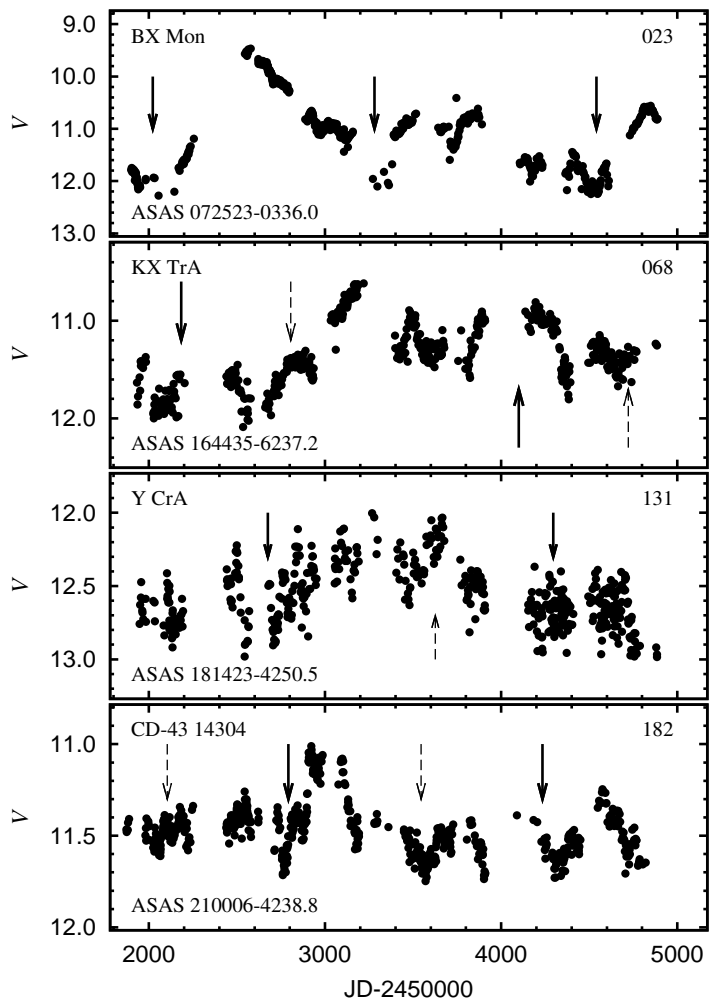}
\end{center}
\caption{Examples of long-term variability in ASAS light curves of BX~Mon, KX~TrA, Y~CrA and CD-43~14304. Arrows show moments of spectroscopic conjunctions (solid: inferior, and dashed: superior).
}
\end{figure}

The ASAS light curves of 8 systems: V417~Cen, HD~330036, Hen~3-1591, Y~CrA, SS73~122,  HD~319167, NSV~11776, and StHA~180 show a wave-like modulation with periods of $\gtrsim$ 1500 days (see examples in Fig. 12).  The first three of them are yellow D'-type systems. In the case of Y~CrA, the similar period ($P$=1619 days) is present in the radial velocity curve of the cool giant, and the orbital solution (Fekel \etal 2010) indicates that the reflection effect may be responsible for optical light modulation (see Fig. 11). Spectroscopic observations of the remaining objects are needed to confirm whether these changes are caused by orbital motion or are due to some other reasons.  

\begin{figure}[htp]
\begin{center}
\includegraphics[width=11.5cm]{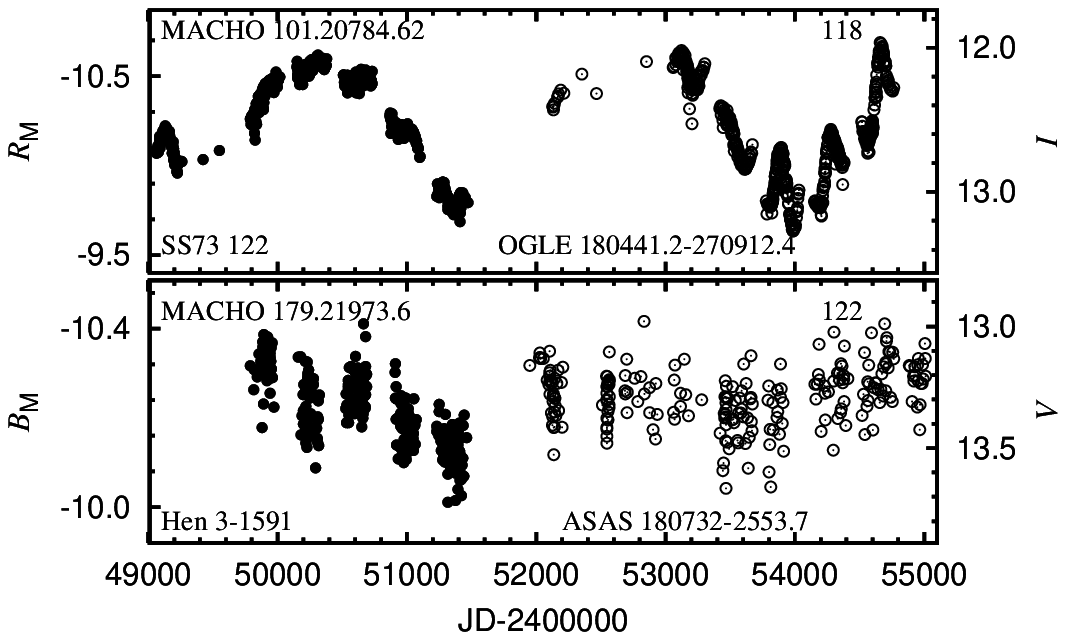}
\end{center}
\caption{Examples of light curves showing modulation with long periods ($\gtrsim$1500 days).}
\end{figure}

The distribution of orbital periods was recently discussed by Miko{\l}ajewska (2012). Above all, although the number of measured periods is continuously increasing ({\it e.g.} Miszalski, Miko{\l}ajewska \& Udalski 2013 discovered 20 new symbiotic systems and found orbital period for 5 S-type systems), the main characteristics of their distribution remain practically the same as in earlier studies ({\it e.g.} Miko{\l}ajewska 2004, 2007). 
At the moment, simulations of the distribution of symbiotic stars over orbital periods with the population synthesis method (PSM) fail to reproduce the observed orbital period distribution of S-type symbiotic binaries. In particular, PSM produces the orbital period distribution in the range 200-6000 days with a maximum at $\approx 1500$ days, and up to 20 \%  of objects with periods below 1000 days (L{\"u} \etal 2006) whereas the observed periods peak at $\approx 600$ days, and only $\approx 30\, \%$ systems have the orbital periods above 1000 days.
This inconsistency cannot be accounted for by selection effects as suggested \eg by L{\"u} \etal (2006). 
At present, 87 systems have known orbital periods, which is about 54 \% of the S-type symbiotic stars included in Belczy{\'n}ski \etal (2000) and Miszalski, Miko{\l}ajewska \& Udalski (2013). 
Additionally, the amplitude of orbital modulation strongly depends on the intrinsically variable luminosity of the hot component, {\it e.g.} the amplitude of variation in visual light of RS~Oph ($i\approx50^{\circ}$, Brandi \etal 2009) varies from $\approx 0.2$ to 0.7 mag (Gromadzki \etal 2008). Assuming as an $i\approx40^{\circ}$ minimum orbital inclination, and the random distribution of orbital inclination angles, we estimate that we should be able to measure the orbital periods for $\approx65$\% of S-type symbiotic systems. 
Then, we already know about 83\% of measurable orbital periods and their distribution cannot be affected by any selection effects.
Successful explanation of the origin of the orbital period distribution of S-type symbiotic binaries requires more advanced approach to mass transfer in these systems, and actually in any interacting binaries involving red giants ( {\it e.g.} Posiadlowski \& Mohamed 2007; Miko{\l}ajewska 2012).

Finally, in systems with eccentric orbits, brightening due to enhanced accretion rate near periastron can be observed regardless of the inclination. Such behaviour is observed in MWC~560 where orbital plane nearly coincides with the plane of the sky, as well as in BX~Mon, which is an eclipsing binary. Symbiotic stars with longer orbital periods, $\gtrsim$1000 days, tend to have eccentric orbits. Observing this kind of variability seems to be very promising and efficient way of deriving their orbital periods.

\subsection{Pulsation periods}

Light curves of 46 objects show, in addition to the long-term or/and orbital variations, short-term variations with time scales of 50-200 days most likely due to stellar pulsations of the cool giant component of the binary, what suggests that the red giants in these systems can be Semi-regular Variables (SRV), or OGLE Small Amplitude Red Giants (OSARG).

The semi-regular red giants are divided into two subtypes: SRa and SRb.
Their basic properties and evolutionary status is described in detail in Kerschbaum \& Hron (1992,1994,1996).
In particular, they found that the SRa appear as intermediate objects between Miras and SRb in all aspects, including periods, amplitudes and mass loss rates. They also concluded that the SRa do not form a distinct class of variables, but are a mixture of 'intrinsic' Miras and SRb. The SRb split into a 'blue' group with $P$ < 150 days and no indication of circumstellar shells and a 'red' group with temperatures and mass loss rates comparable to Miras, but periods about half as long. They suggested that the 'red' and 'Mira' SRb are thermally pulsing AGB stars (Kerschbaum \& Hron 1992). Wood \etal (1999) showed that SRV may obey the same P--L relation (sequence C) as Miras. They are located at the C and C' sequences in the P--L diagram and pulsate in the fundamental mode and in the first overtone, respectively. The mass loss rate of these variables is around $10^{-7}\,M_ {\odot} \, \mathrm{yr}^{-1}$ (Olofsson \etal 2002).

OSARG were first distinguished by Wray \etal (2004) in the Galactic bulge. They found $\approx$18000 red objects, which show pulsation periods with $10 <P_{\rm puls} <$ 100 and the amplitude in the filter $I$ from 0.005 to 0.13. These objects obey different P--L relation than Miras and SRV.
They are on A and B sequences, which are split into ${\rm a}_{1}$, ${\rm a}_{2}$, ${\rm a}_{3}$, ${\rm a}_{4}$, and ${\rm b}_{1}$, ${\rm b}_{2}$, ${\rm b}_{3}$ by Soszy{\'n}ski \etal (2007). They pulsate in the radial modes, indexes represent the order of pulsation mode. Letter ''a'' means AGB objects, ''b'' means RGB objects. 
The P--L relations of ${\rm a}_{\rm k}$ (k=1,2,3,4) sequences extend above the tip of the red giant branch (TRGB), what means that objects located on these sequences are AGB stars. Whereas, the P--L relations of ${\rm b}_{\rm k}$ (k=1,2,3) sequences break off below TRGB, what means that objects located on these sequences are RGB stars (Soszy{\'n}ski \etal 2007). Pulsation periods of ${\rm b}_{\rm k}$ (k=1,2,3) objects are shorter than 70 days. Currently OSARG are the most common type of variable stars. Their number in the LMC, SMC and Galactic bulge 
 is close 300,000 (Soszy{\'n}ski \etal 2009, 2011, 2013). Unfortunately, mass loss rates in these objects are poorly known. 

It is difficult to determine what type of variables are cool giants in galactic symbiotic systems. Distance to most of them is not precisely estimated and we cannot construct for them proper period-luminosity plot. Sequences ${\rm a}_{1}$ and ${\rm b}_{1}$ blend with C'. Classification based on the amplitude of variation seems rather useless, mainly because we have observations in the $V$ filter and a contribution from the hot component in this band make amplitude smaller. Additionally, some pulsation periods are rather tentative due to quality of light curves and high activity of symbiotic systems. On the other hand, typical amplitude of pulsations of OSARG is smaller than scattering of points in ASAS light curves. Despite these difficulties, there is an argument, which indicates that a significant fraction of red giants in S-type symbiotic systems are AGB stars. Most of studied objects (30) show pulsation periods longer than 70 days, what means that these objects have occupied sequences ${\rm a}_{1}$, ${\rm a}_{2}$, ${\rm a}_{3}$, C or C' and they are AGB stars. They cannot develop dusty shell, like cool components in D-type systems, due to influence of nearby hot component. Objects showing pulsation periods in the range 50-70 days may belong to the ${\rm b}_{1}$ or ${\rm b}_{2}$ sequences, although they could be fainter members of ${\rm a}_{1}$, ${\rm a}_{2}$, ${\rm a}_{3}$, C or C' sequences. Fig.~13 shows the distribution of pulsation periods of cool components in studied symbiotic systems. More detailed investigations are needed to fully understand nature of cool companions in symbiotic systems. 

It is worth mentioning that presence of pulsations has been also observed in five systems in Magellanic Clouds: SMC 1, LMC S147, LMC 1, LMC S63 and LMC N67 (Miko{\l}ajewska 2004, Kahabka 2004, Angeloni \etal 2013, Kato, Miko{\l}ajewska \& Hachisu 2013). In all of these systems, red giants are AGB stars because they are brighter than TRGB and only LMC 1 is classified as D-type, others are classified as S-types.

\begin{figure}
\begin{center}
\includegraphics[angle=0,width=11.5cm]{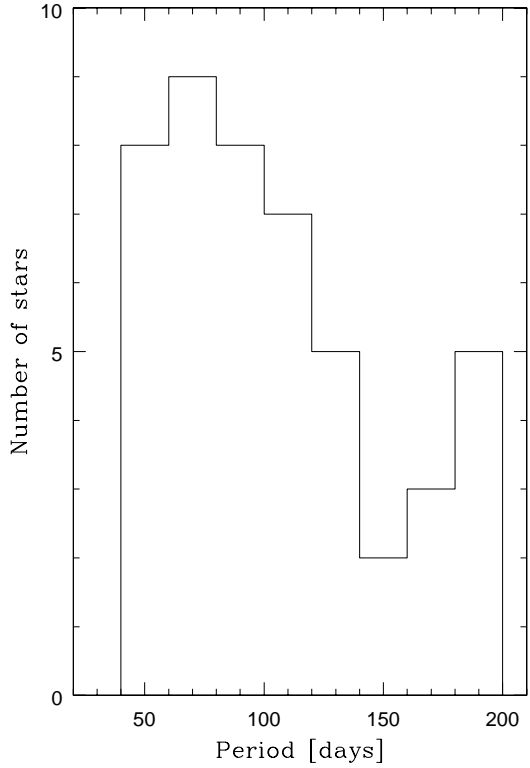}
\end{center}
\caption{Distribution of pulsation periods of cool components in symbiotic binaries.}
\end{figure}

\subsection{Outbursts}

In light curves of 15 systems an outburst is present. Typical duration of such phenomenon is a few years, and the amplitude is from 1 to 3.5 mag in $V$ filter. Such outbursts are common in classical symbiotic stars. Examples of light curves showing outbursts are plotted in Fig. 14. Objects showing outbursts are listed in Table 4.

\begin{figure}[htp]
\begin{center}
\includegraphics[width=11.5cm]{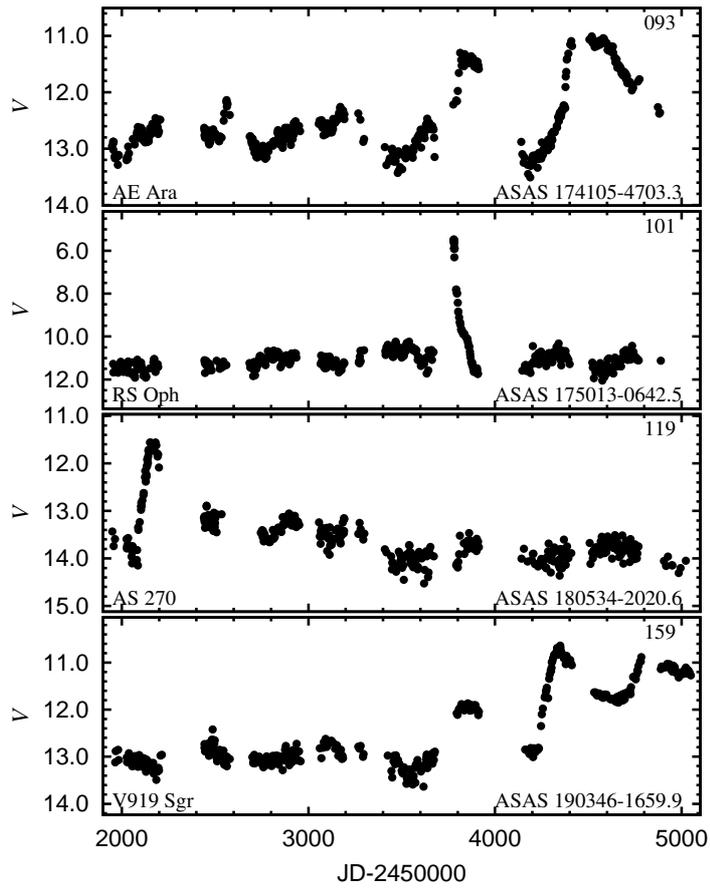}
\end{center}
\caption{Examples of light curves showing outbursts.}
\end{figure}

\begin{table}[hpt]
\caption{Observed outbursts.}	
\label{tab:wyb}
\begin{center}
\begin{footnotesize}
\begin{tabular}{cccc}\hline\hline
No. & Name   & Years & Remarks\\
\hline
035 & RT~Cru       & 1992-2001 & $V_{\rm max}\approx11$ mag \\
068 & KX~TrA       & 2003-2006 & $V_{\rm max}\approx10.5$ \\
070 & HK~Sco       & 2002-2005  & $V\approx13.7-12$ mag   \\
071 & CL~Sco       & 1996-2003, 2009-? & $V\approx13.5-11$ mag \\
074 & Hen~3-1341   & 1998-2003  & $ V\approx13-11$ mag \\
093 & AE~Ara       & 2005-?  &  $V_{\rm max}\approx11$ mag \\
101 & RS~Oph       & 2006 & $V_{\rm max}\approx5$ mag \\
109 & B1~L         & 1998-2000 & $\Delta I~\approx1$ mag \\
119 & AS~270       & 2001 & $V\approx14.5-11.5$ mag  \\
132 & YY~Her       & 2003 & $\Delta V\approx1.5$ mag\\
142 & AR~Pav       & 1984?-2003 &   \\
144 & V3804~Sgr    & 2006 & $\Delta V\approx1$ mag   \\
156 & FN~Sgr       & 2007-? & $\Delta V\approx3$ mag \\
159 & V919~Sgr     & 2005 &  $V\approx13.5-10.6$ mag  \\
\hline\hline
\end{tabular}
\end{footnotesize}
\end{center}
\end{table}

\section{Summary}

In this paper we analysed 79 light curves of S and D'-type symbiotic systems available in ASAS, MACHO and OGLE databases.  The light curves of 58 objects show variations with the orbital period. In most cases (37), these variations are caused by the reflection effect. The remaining objects display ellipsoid modulation and systems with eccentric orbits show brightening related to enhance accretion rate following the periastron passage. 
Eight systems show modulations with period of 1500-2500 days, most probably orbitally related but it may result from instability in accretion, as is the case of RS~Oph (Gromadzki \etal 2008). It is difficult to establish nature of these variations without additional observations.
The orbital periods of 34 S-type symbiotic systems were estimated for the first time, what increases the number of symbiotic stars with known orbital periods by about 64\%. Derived orbital ephemeris cloud be very helpful for planning radial velocities campaigns. 

Light curves of 46 objects show, in addition to the long-term or/and orbital variations, short-term variations with time scales of 50-200 days most likely due to stellar pulsations of the cool giant component of the binary which suggests that the red giants in these systems can be SRV or OSARG. Most of these objects (30) show pulsation periods longer than 70 days, what suggests that they are most likely AGB stars.

Light curves of 15 systems show one or more sharp and deep minima which may be caused by eclipses. Outbursts of hot companions are observed in 15 systems.

\Acknow{This work was partly supported by the Polish Research Grants No. N203\,395534, and DEC-2011/01/B/ST9/06145. MG has been  also financed by the GEMINI-CONICYT Fund, allocated to the project 32110014. 
This study made use of the American Association of Variable 
Star Observer (AAVSO) International Database contributed by observers worldwide and the public domain databases of The All Sky Automated Survey (ASAS) and The Optical Gravitational Lensing Experiment (OGLE), The MACHO Project (MACHO) which we acknowledged. This research has made use of the SIMBAD database, operated at CDS, Strasbourg, France.}

\end{document}